# Statistical thinking in simulation design: a continuing conversation on the balancing intercept problem

Boyi Guo, MS, PhD[1]; Linzi Li, MPH, MSPH[2]; Jacqueline E. Rudolph, PhD[3]

1. Department of Biostatistics, Bloomberg School of Public Health, Johns Hopkins University, Baltimore, MD
2. Department of Epidemiology, Rollins School of Public Health, Emory University, Atlanta, GA
3. Department of Epidemiology, Bloomberg School of Public Health, Johns Hopkins University, Baltimore, MD

**Corresponding Author**:
Boyi Guo, PhD, MS
Department of Biostatistics
Bloomberg School of Public Health
Johns Hopkins University, Baltimore, MD
bguo6@jhu.edu

# Abstract

Epidemiologists have a growing interest in employing computational approaches to solve analytic problems, with simulation being arguably the most accessible among all approaches. While previous literature discussed the utility of simulation and demonstrated how to carry out them, few have focused on connecting underlying statistical concepts to these simulation approaches, creating gaps between theory and application. Based on the recent series of discussions on the *balancing intercept*, we explain the growing complexity when generalizing the balancing intercept to a wider class of simulations and revise the closed-form equation for the balancing intercept under assumptions. The discussion can broadly inform the future design of more complex simulations and emphasize the importance of applying statistical thinking in the new era of computational science.

Epidemiologists have a growing interest in employing computational approaches to solve analytic problems, with simulation being arguably the most accessible among all approaches. Previous papers have argued the importance of simulation in epidemiology education and research [1, 2]. While these papers discussed the utility of simulation and demonstrated how to carry out them, few have focused on connecting underlying statistical concepts to these simulation approaches, creating gaps between theory and application. Here, we seek to put commonly used statistical concepts, including variable enumeration, generalized linear model, and link functions, in the context of simulation methods. Based on the recent series of discussions on the balancing intercept [3], we explain the growing complexity when generalizing the balancing intercept to a wider class of simulations and revise the closed-form equation for the balancing intercept under assumptions. The discussion can be broadly helpful to the understanding of more complex simulation designs, even in the context of causal inference settings and emphasize the importance of applying statistical thinking in the new era of computational science.

## REVIEWING THE BALANCING INTERCEPT

The balancing intercept, first introduced by Rudolph et al. [3], is an estimation of the unknown intercept in regression-based data-generating mechanisms to control the marginal mean of a simulated variable. To explain with a toy example, suppose we are interested in simulating a normally distributed outcome ($Y$) conditioning on a binary exposure ($X$) with known group sizes. Our goal is to parameterize the simulation using the mean difference between the exposure groups ($\beta_1$), commonly reported in the literature, such that the marginal mean $E(Y)$ is fixed at a level of interest. Without the group means $E(Y|X)$ directly available, one needs to calculate the intercept ($\beta_0$) in a regression system $E(Y|X) = \beta_0 + \beta_1 X$. Acknowledging the degree of freedom is fixed, Rudolph et al. (2021) provided a closed-form equation to calculate the $\beta_0$, referred to as the balancing intercept,

$$E(Y) = E(E(Y|X)) = \beta_0 + \beta_1 E(X) \ \rightarrow \ \beta_0 = E(Y) - \beta_1 E(X). \qquad (1)$$

Obviously, most, if not all, simulation designs are more complex than a two-sample normal outcome design. For example, we often consider various outcome types (e.g., categorical, survival, or other, complex continuous distributions), continuous or multinomial exposures (i.e. exposures with more than 2 levels), and different estimands of interest. Mirroring real-world observational data, we also routinely need to adjust for cofounders. Equation (1) does not generalize to these complex designs, as was first noticed by Robertson et al [4].

## CONNECTING SIMULATION TO REGRESSION FUNDAMENTALS

Complex simulation designs require specifying a data-generating model carefully, normally written as a series of structural equations [2]. For simplicity, those equations are assumed to follow some parametric forms and are often expressed as a generalized linear model (GLM). Hence, familiarizing with the concepts of GLM can address challenges when deriving the balancing intercept in more complex data-generating models.

## Outcome and Estimand

Given the type of outcome, data are sampled from a distribution, such as Gaussian distribution for continuous outcomes, Bernoulli distribution for binary outcomes, and Weibull distribution for survival time. In addition, a link function, $f(\cdot)$, that describes the expected mathematical relationship between the exposure and the mean of the outcome is specified in the simulation design. The choice of the link function is highly relevant to the type of outcome and dictated by the estimand of interest. Even though the logit function is commonly the default link function to simulate binary outcomes thanks to its mathematical property of being bounded, it is still useful and possible to consider other link functions when investing different estimands, e.g. a logarithmic (log) function to study the risk ratio, a logit function to study the odds ratio, or an identify function to study the risk difference. We refer to the risk ratio, odds ratio, or risk difference as the estimand of interest.

Different choices of the link function lead to different complexities to derive the balancing intercept, with a linear function being the easiest to calculate (see Equation (1)). When the link function is nonlinear, e.g. log function, the equality between the expectation of a link function and the link function of an expectation no longer holds, $E(f(Y)) \neq f(E(Y))$, as shown in Jensens' inequality [5]. Consequently, Equation (1) does not hold for non-linear link functions. See the Supporting Information for the complete mathematical reasoning.

Is it possible to derive a closed-form equation for non-linear link functions? Yes, but only for a limited number of link functions. When applying a log link function, the balancing intercept is calculated (derivations shown in Supporting Information) via

$$\beta_0 = \log E(Y) - \log E[\exp(\beta_1 X)]. \qquad (2)$$

In contrast, the logit function, another popular link function in simulation designs, does not have a tractable solution. Hence, we recommend using numeric approximation approaches, discussed in Robertson et al. [4] and Zivich and Ross [6].

## Covariates

Statistical covariates are one of the most fundamental concepts in data analysis. The covariates could be confounders that affect both the exposure and the outcome, or mediators and effect measure modifiers that affect the outcome. In simulation designs, the multivariate distribution of exposure and covariates needs to be accounted for, which increases the complexity of the balancing intercept calculation.

When the exposure $X$ and the covariates $\mathbf{Z}$, are pairwise independent, the joint mean $E_{X,Z}$ can be written as a function of marginal means $E_X$ and $E_Z$. For example, when applying to simulation designs with the log link functions, Equation (2) extends as

$$\beta_0 = \log E(Y) - \log E[\exp(\beta_1 X)] - \sum_{i=1}^{p} \log E[\exp(\beta_{i+1} Z_i)]. \qquad (3)$$

We can simplify this calculation by replacing the expectation of the exponential function, $E(\exp(\beta_1 X))$, with its moment generating function [7]. Similar to probability distribution functions, the moment generating function uniquely characterizes a data distribution and provides a shortcut to the statistical moments of the distribution. For example, the moment generating function for normally distributed $X$ with mean $\mu$ and variance $\sigma^2$ is $M_\beta(X) = E(\exp(\beta X)) = \exp(\mu\beta + \sigma^2\beta^2/2)$.

Confounding and effect mediation set up hierarchical structures in the simulation of statistical covariates and undermine the pairwise independent assumption. In addition, not all distributions have moment generating functions, e.g. the Cauchy distribution. In these situations, we can apply the Monte Carlo technique to derive $E[\exp(\beta_1 X + \beta_2 Z)]$. Specifically, one can sample the vector of variables with replacement for a large number of iterations (say 1000) and average the exponential function of the randomly sampled data.

### Multinomial covariates and Coding Schemes

Does having a multinomial exposure, in contrast to a binary exposure, complicate the calculating balancing intercept? Surprisingly, the short answer is no, due to its discrete nature. When enumerating a nominal variable with $p$ levels, we normally compose a data matrix (denoted as $X$) where each column represents one level of this nominal variable. The expectation $E(f(\boldsymbol{\beta}^T \mathbf{X}))$ for the nominal variable with the sampling probabilities ($\boldsymbol{\pi} = \{\pi_1, \ldots, \pi_p\}$) with the corresponding effect coefficients ($\boldsymbol{\beta} = \{\beta_1, \ldots, \beta_p\}$) on the appropriate scale (link function), $f(\cdot)$, can be expressed as $E(f(\boldsymbol{\beta}^T \mathbf{X})) = \sum_{i=1}^{p} \pi_i f(\boldsymbol{\beta}^T X_i)$. In the presence of statistical interactions, one can simply treat the statistical interaction as a special case of a multinomial variable by enlisting all possible combinations.

The enumeration of each level ($X_i, i = 1, \ldots, p$) in this multi-column matrix is called a coding scheme. The most popular coding scheme, as well as the default in the previous balancing intercept literature, is the reference cell coding (also known as dummy coding). With the reference cell coding, the data matrix consists of $p - 1$ column to indicate these $p$ levels, assuming the reference level collapse with the intercept term. Each column marks the membership in a corresponding level using either 1 or 0. The expectation can be simply written as $E(f(\boldsymbol{\beta}^T \mathbf{X})) = \sum_{i=1}^{p} \pi_i f(\beta_i)$ with $\beta_1 = 0$. There exist other coding schemes that can possibly simplify the calculation of the balancing intercept. For example, effect coding encodes different levels with a combination of 0, 1, and -1, with the coefficients emphasizing the deviation from the grand mean, i.e. the average of level means. When a study is balanced with respect to the exposure variable, i.e. having equal number of observations in each exposure group, the grand mean coincides with the marginal mean. Hence, the balancing intercept is the targeted marginal mean under the effect coding scheme, requiring no further calculation. In the case of an unbalanced design, one can replace effect coding with the weighted effect coding [8].

## SIMULATION EXAMPLE

To demonstrate the closed-form equation (Equation (3)), we conduct a Monte Carlo simulation study motivated by Robertson et al. [4]. The simulation follows a log-normal model with two variables that are statistically independent, an exposure ($X$) and a covariate ($Z$). We assumed $X$ was a three-level categorical variable with the sampling probabilities (0.5, 0.35, 0.15). We examined different distributions for $Z$: (1) a Bernoulli distribution with probability

0.8, (2) a continuous uniform distribution bounded between -1 and 3, (3) a standard normal distribution, and (4) a gamma distribution with shape 1 and rate 1.5. We also examined different magnitudes of covariate coefficient $\beta_2$ ranging from 1 to 3 with 0.5 increments, while fixing the coefficients $\boldsymbol{\beta_1}$ for the exposure $X$ at (0.2, -0.2). We evaluated multiple targeted marginal means $E(Y) = exp(\beta_0 + \boldsymbol{\beta_1} X + \beta_2 Z)$, ranging from 0.1 to 0.9 with 0.1 increments, where $Y$ is sampled from a normal distribution with standard deviation 0.1. For each combination of these parameters, we used Equation (3) to calculate the balancing intercept and simulated a dataset of 10,000 observations. We calculated the deviation of the observed mean from the target mean, referred to as bias. We iterated the process 10,000 times and calculated the average bias. Figure 1 shows that the closed-form equation produced unbiased estimates of $E(Y)$ in the simulated sample.

We also examined the bias when applying Equation (3) to a binomial outcome $Y$, instead of a continuous outcome, while keeping the log link function. (See supporting information Figure 1) We observed that it is difficult to control the marginal mean with an analytic solution of balancing intercept, particularly when the effect size is large. This can be explained by the fact that the outcome is bounded, e.g. probabilities or binary outcomes, while the link function is not, creating asymmetry that skews the distribution of the mean.

## CONCLUSIONS

In this commentary, we highlighted how standard simulation approaches rely on the fundamentals of statistics and parametric regression, such as link function, expectation calculation, moment generating functions, and the coding scheme. We described how to extend the balancing intercept for various link functions, the inclusion of covariates, and the generalization to multinomial variables. We also derived a close-form equation to calculate the balancing intercept for simulation designs with the log link function. Simulation studies were conducted to demonstrate that the close-form equation produced unbiased estimates of the marginal mean of the outcome.

When introduced in the statistical training required by most epidemiology programs, statistical concepts like coding schemes and moment-generating functions can appear merely theoretical or academic – something to be learned for a test but ultimately never used in practice. However, these concepts remain useful in many settings, including simulation. Using the balancing intercept problem as an example, we showed how these statistical concepts can address computational challenges in real-world problems. The balancing intercept problem provides a didactic platform to exemplify these concepts and a case study that facilitates students to understand elements of data-generating models. This case study reminds us of the importance of fundamental statistics training in epidemiology, even in the new era of computation.

# Reference

1. Rudolph, J.E., M.P. Fox, and A.I. Naimi, *Simulation as a Tool for Teaching and Learning Epidemiologic Methods.* Am J Epidemiol, 2021. **190**(5): p. 900-907.
2. Fox, M.P., et al., *Illustrating How to Simulate Data From Directed Acyclic Graphs to Understand Epidemiologic Concepts.* Am J Epidemiol, 2022. **191**(7): p. 1300-1306.
3. Rudolph, J.E., et al., *SIMULATION IN PRACTICE: THE BALANCING INTERCEPT.* American Journal of Epidemiology, 2021. **190**(8): p. 1696-1698.
4. Robertson, S.E., J.A. Steingrimsson, and I.J. Dahabreh, *Using Numerical Methods to Design Simulations: Revisiting the Balancing Intercept.* American Journal of Epidemiology, 2022. **191**(7): p. 1283-1289.
5. Durrett, R., *Probability: theory and examples*. Vol. 49. 2019: Cambridge university press.
6. Zivich, P.N. and R.K. Ross, *RE: "USING NUMERICAL METHODS TO DESIGN SIMULATIONS: REVISITING THE BALANCING INTERCEPT".* American Journal of Epidemiology, 2022.
7. Casella, G. and R.L. Berger, *Statistical inference*. 2021: Cengage Learning.
8. Te Grotenhuis, M., et al., *When size matters: advantages of weighted effect coding in observational studies.* International Journal of Public Health, 2017. **62**(1): p. 163-167.

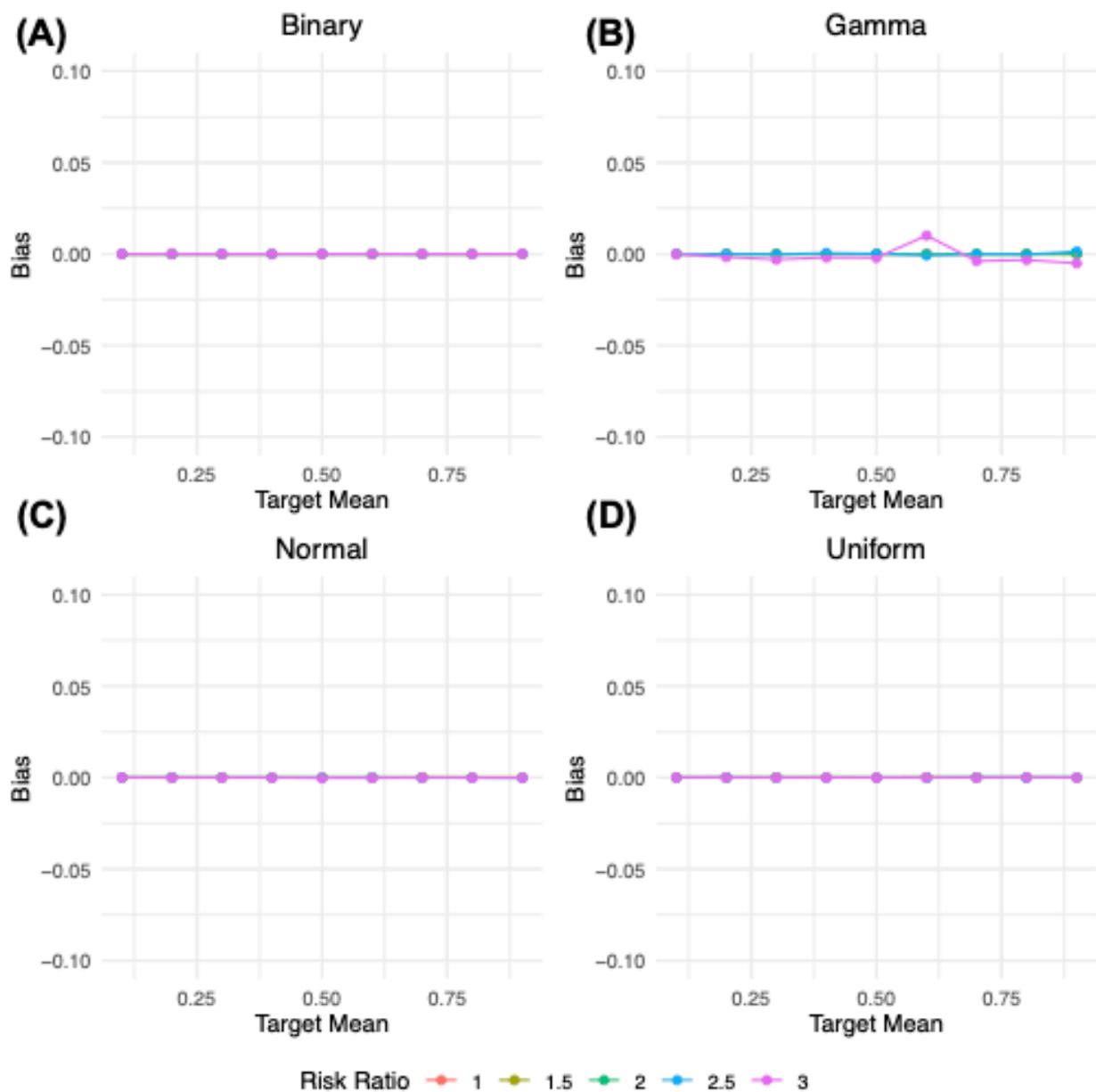

**Figure 1**: **The closed-form estimation of the balancing intercept controls the marginal mean at the target level for log-normal data generating models with different covariates settings.** The bias, defined as the empirical mean of the simulated outcome minus the targeted marginal mean of the outcome, holds at 0 for log-normal data generating models of four different risk ratio magnitude of the covariates and four different distributions for the covariates, including (**A**) a Bernoulli distribution with probability 0.8, (**B**) a gamma distribution with shape 1 and rate 1.5, (**C**) a standard normal distribution, and (**D**) a continuous uniform distribution bounded between -1 and 3.

## Ethics approval and consent to participate
Not applicable.

## Competing interests
The authors declare that they have no competing interests.

## Conflict of Interest
None declared.

# Supporting Information

Statistical thinking in simulation design: a continuing conversation on the balancing intercept problem

Boyi Guo, Linzi Li, Jacqueline E. Rudolph

The marginal mean (probability of event for binary outcome), $\mathbb{E}_Y(Y)$ can be expressed as a double expectation of the covariates $\boldsymbol{X}$

$$\begin{aligned}\mathbb{E}_Y(Y) &= \mathbb{E}_{\boldsymbol{X}}(\mathbb{E}_Y(Y|\boldsymbol{X})) \\ &= \mathbb{E}_{\boldsymbol{X}}(g^{-1}(\beta_0 + \boldsymbol{\beta}_1 \boldsymbol{X})),\end{aligned}$$

where $g^{-1}$ is the inverse function of the link function $g$, $\beta_0$ is the balance intercept of interest, $\boldsymbol{\beta}_1$ is the coefficient vector for the covariates $\boldsymbol{X}$ which can be the enumeration of a categorical variable or multiple continuous variable.

**Balance Intercept Calculation on the Linear Predictor Scale**

Only when $g^{-1}(\mathbb{E}(\cdot)) = \mathbb{E}(g^{-1}(\cdot))$ (e.g. $g^{-1}$ is a linear function), we can accurately calculate the balance intercept on the linear predictor scale, as

$$\begin{aligned}\mathbb{E}_Y(Y) &= g^{-1}\{\mathbb{E}_{\boldsymbol{X}}(\beta_0 + \boldsymbol{\beta}_1 \boldsymbol{X})\} \\ g\{\mathbb{E}_Y(Y)\} &= \mathbb{E}_{\boldsymbol{X}}(\beta_0 + \boldsymbol{\beta}_1 \boldsymbol{X}) \\ g\{\mathbb{E}_Y(Y)\} &= \beta_0 + \mathbb{E}_{\boldsymbol{X}}(\boldsymbol{\beta}_1 \boldsymbol{X}) \\ \beta_0 &= g\{\mathbb{E}_Y(Y)\} - \mathbb{E}_{\boldsymbol{X}}(\boldsymbol{\beta}_1 \boldsymbol{X}).\end{aligned} \quad (1)$$

Equation (1) further simplifies to the analytic approximation in Rudolph et al. (2021) when $\boldsymbol{X}$ are pairwise independent,

$$\beta_0 = g\{\mathbb{E}_Y(Y)\} - \sum_{j=1}^{p} \beta_j \mathbb{E}_{X_j}(X_j).$$

**Balance Intercept Calculation on the Response Scale**

When $g^{-1}(\mathbb{E}(\cdot)) \neq \mathbb{E}(g^{-1}(\cdot))$, it would be only sensible to conduct the calculation on the response scale. And the calculation can be complicated as the complexity of the link function and the number of predictors increase. We show how to derive the balance intercept when the link function is the logarithm function, i.e. $g(x) = \log(x)$.

$$\begin{aligned}\mathbb{E}_Y(Y) &= \mathbb{E}_{\boldsymbol{X}}(\exp(\beta_0 + \boldsymbol{\beta}_1 \boldsymbol{X})) \\ &= \mathbb{E}_{\boldsymbol{X}}(\exp(\beta_0) * \exp(\boldsymbol{\beta}_1 \boldsymbol{X})) \\ &= \exp(\beta_0) \mathbb{E}_{\boldsymbol{X}}(\exp(\boldsymbol{\beta}_1 \boldsymbol{X})) \\ \exp(\beta_0) &= \mathbb{E}_Y(Y)/\mathbb{E}_{\boldsymbol{X}}(\exp(\boldsymbol{\beta}_1 \boldsymbol{X}) \\ \beta_0 &= \log\{\mathbb{E}_Y(Y)/\mathbb{E}_{\boldsymbol{X}}(\exp(\boldsymbol{\beta}_1 \boldsymbol{X})\} \\ \beta_0 &= \log(\mathbb{E}_Y(Y)) - \log(\mathbb{E}_{\boldsymbol{X}}(\exp(\boldsymbol{\beta}_1 \boldsymbol{X}))).\end{aligned} \quad (2)$$

When $\boldsymbol{X}$ are pairwise independent, Equation (2) simplifies to

$$\beta_0 = \log\{\mathbb{E}_Y(Y)\} - \sum_{j=1}^{p} \log(\mathbb{E}_{X_j}(\exp(\beta_j X_j))).$$

If the moment generating function $M_X(t)$ is known, it can be used to simplify the calculation for $M_X(\beta) = \mathbb{E}_X(\exp(\beta X))$ for individual variables or $M_{\boldsymbol{X}}(\boldsymbol{\beta}_1) = \mathbb{E}_{\boldsymbol{X}}(\exp(\boldsymbol{\beta}_1 \boldsymbol{X})$ for the variable vector.

For some more complicated link function, e.g. the logistic function $g(x) = log\frac{x}{1-x}$, or the distribution of covariate $X$ is unknown, it is very difficult, if not impossible, to derive the closed-form solution of the balance intercept. Hence, it would be preferred to use the numeric solution.

**Supplementary Figure 1**

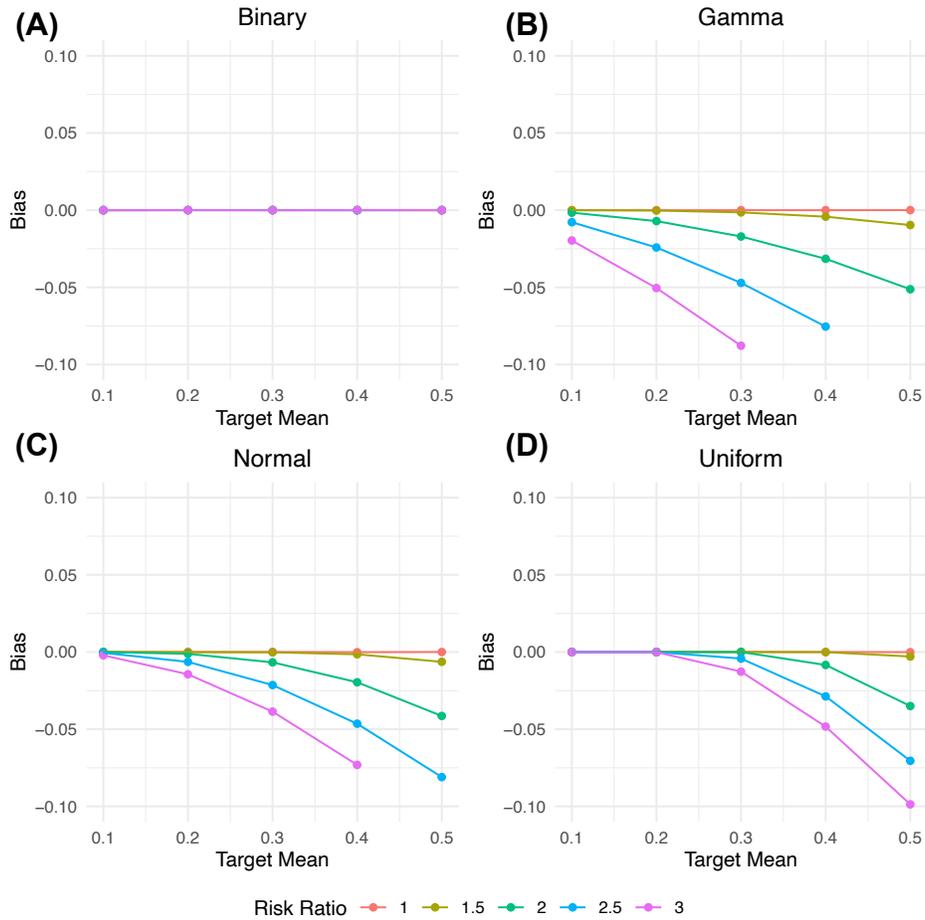

Figure 1: **Unbound link functions for bounded outcomes can be diffcult to control the marginal mean using a closed-form solution**: The bias, defined as the empirical mean of the simulated outcome minus the targeted marginal mean of the outcome, holds at 0 for log-binoimal data generating models of four different risk ratio magnitude of the covaraites and four different distribution for the covariates, including (**A**) a Bernoulli distribution with probability 0.8, (**B**) a gamma distribution with shape 1 and rate 1.5, (**C**) a standard normal distribution, and (**D**) a continuous uniform distribution bounded between -1 and 3.